\pdfoutput=1
\documentclass[amsmath,amssymb,aps,prl,twocolumn,superscriptaddress,floatfix]{revtex4-2}
\usepackage{graphicx}
\usepackage{wrapfig}
\usepackage[usenames]{color}
\usepackage[bookmarks=false,linkcolor=NavyBlue,urlcolor=NavyBlue,colorlinks,citecolor=NavyBlue]{hyperref}
\usepackage[dvipsnames]{xcolor}

\newcommand{\figref}[2]{\hyperref[#1]{\ref{#1}(#2)}}
\newcommand{\figsref}[2]{\hyperref[#1]{\ref{#1}#2}}

\renewcommand{\vec}[1]{\mathbf{#1}}

\usepackage[normalem]{ulem}

\newcommand{\xCornell}{Department of Physics, Cornell University, Ithaca, NY 14853, USA}

\newcommand{\xEwha}{Department of Physics, Ewha Womans University, Seoul, South Korea\\\vspace{0.7em}}
\newcommand{\xKentucky}{Department of Physics and Astronomy, University of Kentucky, Lexington, Kentucky 40506-0055, USA}

\begin{document}

\title{Emblems of pair density waves: dual identity of topological defects and their transport signatures}

\author{Omri Lesser}
\affiliation{\xCornell}
\author{Chunli Huang}
\affiliation{\xKentucky}
\author{James P. Sethna}
\affiliation{\xCornell}
\author{Eun-Ah Kim}
\affiliation{\xCornell}
\affiliation{\xEwha}

\begin{abstract}
The pair density wave (PDW) exemplifies intertwined orders in strongly correlated systems.  A recent discovery of superconductivity in a quarter-metal state~\cite{han_signatures_2025} offers the first experimental system where a pure PDW without uniform superconductivity is suspected, offering a unique opportunity to examine the consequences of intertwined orders.  A pure two-dimensional PDW supports an unusual fractional excitation as its topological defect (TD). A TD simultaneously winds the phase of the Cooper pair and distorts the amplitude modulation—a dual role reflecting its intertwined character.  As a vortex, a TD carries fractional vorticity of $\frac{1}{3} h/2e$, whose movement would cause resistance. As a crystalline defect, a TD can be sourced by charge disorder in the system. We show that experimentally observed resistive switching can originate from mobile TDs, while a small magnetic field will restore zero resistance by blocking their motion. The resulting resistive state exhibits extreme anisotropy and a Hall response, with the Hall angle determined by the angle between the current and the TD's Burgers vector. These features will serve as confirmation of the dual identity of topological defects as emblems of PDW order.
\end{abstract}
\maketitle

Ever since intriguing transport observations in cuprate superconductors invited the invocation of the concept~\cite{berg_charge-4e_2009}, the notion of a pair density wave (PDW), a finite-momentum carrying paired state, has become a fixture in studies of strongly correlated superconductors such as cuprates~\cite{edkins_magnetic_2019,choubey_atomic-scale_2020,wang_scattering_2021,dai_pair-density_2018-1}, UTe$_{2}$~\cite{gu_detection_2023,aishwarya_melting_2024}, and kagome materials~\cite{li_discovery_2022-1,deng_chiral_2024,wu_sublattice_2023-1,schwemmer_sublattice_2024,yao_self-consistent_2025}. However, these systems all contain charge density wave order and a uniform superconducting component. 
The fact that the modulated component's amplitude is subdominant incurred the criticism of the PDW being a subsidiary effect. Practically, such subdominance also meant many of the interesting properties of PDWs have been hard to resolve. 

\begin{figure}[ht!]
    \centering
    \includegraphics[width=\linewidth]{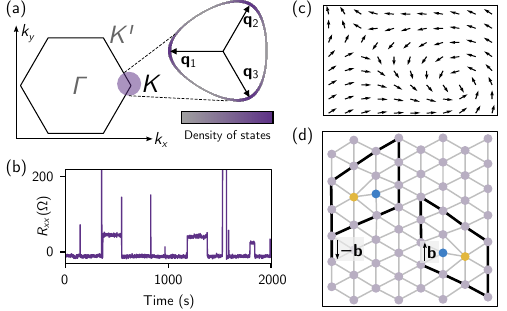}
    \caption{(a)~The Brillouin zone of (multilayer) graphene. In the quarter metal phase, only one of the valleys ($K$ in this illustration) is occupied. The Fermi surface around $K$ is trigonally warped, leading to three hotspots with high density of states, denoted by the momenta $\vec{q}_1$, $\vec{q}_2$, $\vec{q}_3$.
    (b)~Experimental data from Ref.~\cite{han_signatures_2025} showing time-dependent resistance fluctuations in one of the superconducting states of rhombohedral tetralayer graphene.
    (c)~Illustration of the phase of a ordinary two-dimensional superconductor with a vortex and an antivortex.
    (d)~5 (blue) -7 (yellow) pair crystalline defect of a triangular lattice. The loops around the defects show the Burgers vector $\vec{b}$.
    }
    \label{fig:general_picture}
\end{figure}

An intriguing recent experiment on rhombohedral tetralayer graphene~\cite{han_signatures_2025} invites us to think about pure PDW order. In the experiment, superconductivity was reported in a two-dimensional material with the parent state being a spin- and valley-polarized quarter metal; see Fig.~\figref{fig:general_picture}{a}. In such a state, superconducting pairing must take place \emph{within} a valley, leading to finite Cooper pair momentum: pure PDW, without a uniform component, is natural~\cite{han_pair_2022}. Several theoretical works have considered the possible mechanisms and order parameter symmetries in this unusual superconductor~\cite{chou_intravalley_2025,qin_chiral_2024,yang_topological_2024,geier_chiral_2024,kim_topological_2025,yoon_quarter_2025,parra-martinez_band_2025,daido_current-induced_2025,gil_charge_2025,christos_finite-momentum_2025,jahin_spontaneous_2025}. However, the connection between the observations and the PDW nature remains unclear.

The phenomenology observed in Ref.~\cite{han_signatures_2025} is mostly in line with conventional superconductivity, except for one significant departure depicted in Fig.~\figref{fig:general_picture}{b} (adapted from Ref.~\cite{han_signatures_2025}): the zero-field resistance exhibits temporal switching behavior (telegraph noise)~\cite{ralls_individual-defect_1989}. This behavior, unique to the phase labeled SC1 in Ref.~\cite{han_signatures_2025}, is the main subject of this paper, as a potential harbinger of the pair density wave order.
In what follows, we will argue that topological defects in the PDW order can explain the switching behavior shown in Fig.~\figref{fig:general_picture}{b}. The key realization is that defects in a multi-component PDW have a dual identity: the Cooper pair phase winds around a topological defect (TD) [see Fig.~\figref{fig:general_picture}{c}], which is simultaneously a 5-7 pair crystalline dislocation [see Fig.~\figref{fig:general_picture}{d}].
The physics of vortices and that of crystalline dislocations are individually well-studied. 
The proliferation of vortices in two-dimensional superconductors leads to the destruction of quasi long-range order via the Berezinskii-Kosterlitz-Thouless mechanism~\cite{kosterlitz_ordering_1973-1,jos_40_2013}; and in two-dimensional crystals, the proliferation of structural defects leads to  melting~\cite{nelson_study_1978,nelson_dislocation-mediated_1979,hayashi_effects_2002}. 
Here we study the formation of TDs in pairs at nucleation centers -- defects or soft spots in the superconducting order (see Fig.~\ref{fig:defects_motion}).
Nucleation is dominated by defects (e.g., dust grains nucleating raindrops in water vapor stressed by supercooling); nucleation of TDs at defects allows for local release of the stresses (e.g., Frank-Read sources~\cite{frank_multiplication_1950} nucleating slip through the release of a periodic array of 3D dislocations under shear stress~\cite{nicolas_deformation_2018,baggioli_plasticity_2021,wu_topology_2023}). 
Current will impose stress on the superconducting vortex component of our defects, mediating a continuous stream of TDs causing the observed resistance jumps.

\begin{figure}
    \centering
    \includegraphics[width=\linewidth]{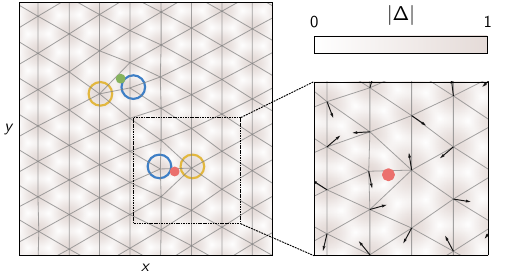}
    \caption{The PDW and its TDs. The colorscale indicates the PDW order parameter's magnitude, whereas the arrows correspond to its phase. The three phase-modulated components lead to both amplitude and phase modulations, with the amplitude maxima defining an emergent triangular lattice. The 5 (yellow) -7 (blue) crystalline defect pairs are TD's with $\pm\frac{1}{3}$ flux quantum attached due to $2\pi$ winding of only one of the three components of the order parameter.
    } 
    \label{fig:lattice_with_defects}
\end{figure}

We begin by describing the long-range ordered PDW pattern without TDs. As indicated in Fig.~\figref{fig:general_picture}{a}, trigonal warping induces three hotspots in the Fermi surfaces, with momenta $\vec{q}_1$, $\vec{q}_2$, $\vec{q}_3$ relative to the $K$ point. The most natural form of spatial modulation is of the FF type (phase modulation). The overall pairing order parameter in this case is 
\begin{equation}\label{eq:pdw_pattern}
    \Delta(\vec{r})=\sum_{j=1}^{3} \Delta_{\vec{q}_{j}} e^{i(2\vec{K}+\vec{q}_{j})\cdot\vec{r}}, 
\end{equation}
where $\{\Delta_{\vec{q}_{j}}\}$ are the three order parameters corresponding to the three modulation vectors $\{\vec{q}_{j}\}$ that are incommensurate with the underlying lattice, leading to $U(1)\times U(1)\times U(1)$ symmetry. In the $C_3$ symmetric case, the $\{\Delta_{\vec{q}_{j}}\}$ are all equal, resulting in the modulation pattern shown in Fig.~\ref{fig:lattice_with_defects}; see also Refs.~\cite{agterberg_conventional_2011,gaggioli_spontaneous_2025}. Due to the three-component structure of the PDW, the pure FF phase modulation now also has a LO (amplitude) component, see Fig.~\ref{fig:lattice_with_defects}, forming a honeycomb lattice of vortices and antivortices. This crystal of Cooper pairs can be viewed as an antiferromagnetic lattice of vortices, and it forms the background upon which we will now study topological defects.  This emergent crystal will be pinned by charge disorder, keeping vortices frozen in place without causing resistance.

As was first noted in~\cite{agterberg_dislocations_2008,berg_charge-4e_2009}, unlike conventional superconductors, unidirectional PDWs with order parameter $\Delta(\vec{r}) = (\Delta_{\vec{Q}},\Delta_{-\vec{Q}}) \propto (e^{i\phi_1}, e^{i\phi_2})$ support half-quantum vortices where a single phase $\phi_1$ or $\phi_2$ winds by $2\pi$, yielding a magnetic flux of $h/4e$. The consistency of these half-quantum vortices with a single-valued wavefunction is ensured by their composite nature, where a $\pi$ phase winding in the superconducting order is accompanied by a half-dislocation in the periodic PDW structure. This is similar to the way half-quantum vortices are allowed in spin-triplet superconductors, where single-valuedness of the superconducting order parameter is guaranteed by phase winding in the spin degree of freedom, making up for half vorticity~\cite{agterberg_dislocations_2008,chung_stability_2007}.  
In our case, the presence of three order parameters leads to an analogous ``$\frac{1}{3}$ vortex" as the lowest-energy defect (see Sec.~SI of the Supplemental Material). Figure~\ref{fig:lattice_with_defects} shows a pair of such fractional vortices, introduced by winding only $\Delta_{1}$ of Eq.~\eqref{eq:pdw_pattern}. 
It is evident that, besides their vortex nature, these defects are also dislocations of the triangular lattice, known as 5-7 pairs~\cite{kosterlitz_ordering_1973-1,nelson_dislocation-mediated_1979,young_melting_1979,pretko_fracton-elasticity_2018,radzihovsky_fractons_2020,gromov_colloquium_2024}: in the pristine triangular lattice each site has 6 neighbors, and the defects distort the lattice such that one site has 5 neighbors and another has 7. 
A unique feature of our three-component PDW is that it enables a magnetic object --- the $\pm\frac{1}{3}$ vortex --- to be sourced and pinned by charge impurities, through its dual role as a crystalline defect. 
This unusual possibility of charge impurities sourcing a fractional vortex-antivortex pair, $\left\{\frac{1}{3},-\frac{1}{3}\right\}$, without an external magnetic field, reflects the crux of the intertwined nature of the PDW.

\begin{figure*}
    \centering
    \includegraphics[width=\linewidth]{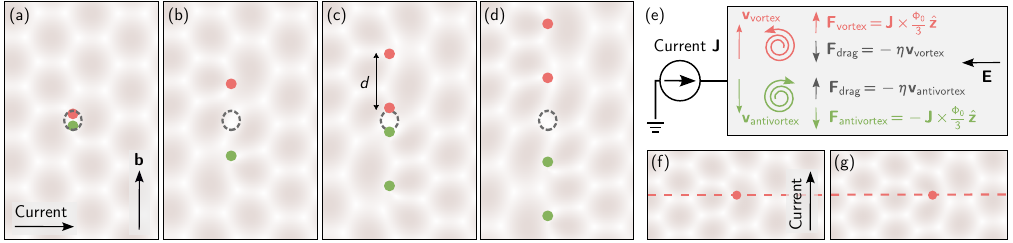}
    \caption{
    The motion of defects in the pair density wave background and its associated resistance. 
    (a)--(b)~A weak spot in the superconductor (dashed circle) becomes a constant source of defect-antidefect pairs. Current makes the defects (red) and antidefects (green) move in opposite directions. The Burgers vector $\vec{b}=b\hat{\vec{y}}$ associated with the defects is vertical.
    (c)--(d)~When the defects get far enough from the source, a second pair of defects leaves it and starts moving with the same velocity. 
    (e)~Bardeen-Stephen mechanism for resistance due to vortex motion. The horizontal current ${\bf J}$ induces a Lorentz force on the $\pm\frac{1}{3}$ vortices, making them move in opposite directions, and the viscosity $\eta$ determines their terminal velocity. The motion of fluxes causes an electric field $\vec{E}\propto -\vec{v}_{\rm vortex}\times\frac{\Phi_0}{3}\hat{\vec{z}}$ in the horizontal direction, which leads to a voltage drop and hence to resistance.
    (f)--(g)~The background landscape of modulated superconductivity makes it harder for a defect to climb in the horizontal direction (dashed line) compared to gliding in the vertical motion shown in panels (a)--(d).
    } 
    \label{fig:defects_motion}
\end{figure*}

Having identified the elementary topological defects, we now discuss their motion and its relation to resistance. Due to the amplitude modulation of the pairing order parameter, a weak spot (caused by a charge impurity) can serve as a source of TD-anti TD pair~\cite{frank_multiplication_1950} at zero magnetic field, as illustrated in Figs.~\figref{fig:defects_motion}{a--d}. When a current $\vec{J}$ is applied in the $\hat{\vec{x}}$ direction, the TDs and anti-TDs move in the $\pm\hat{\vec{y}}$ direction, respectively, due to the Lorentz force acting on them by virtue of their nature as $\pm\frac{1}{3}$ vortices. Within the Bardeen-Stephen theory~\cite{stephen_viscosity_1965}, this force is balanced by a viscous force $-\eta\vec{v}$, leading to a constant terminal velocity of the defects. Since defects of the same topological charge logarithmically repel each other, the TD and anti-TD both must travel some distance $d$ from the charge-impurity source before a new pair of defects is generated. We therefore get a steady flow of defects along the vertical direction, leading to a voltage buildup in the horizontal direction, as depicted in Fig.~\figref{fig:defects_motion}{e}.

Within the Bardeen-Stephen theory~\cite{stephen_viscosity_1965}, the resistance due to a single mobile vortex is $R=\Phi_{0}^{2}/(\eta L_{x}L_{y})$, where $L_x,L_y$ are the physical dimensions of the system and $\Phi_{0}=h/2e$. For a field-induced full vortex whose superconducting order parameter is completely suppressed at its core, the viscosity is phenomenologically estimated to be 
$\eta_{0}=\Phi_{0}^{2}/(2\pi\xi^{2}R_{\rm N})$, where $R_{\rm N}$ is the normal-state resistance and $\xi$ is the SC coherence length (see Sec.~SII of the Supplemental Material).
This result hinges on the assumption that the vortex core is completely metallic with resistance $R_{\rm N}$ and that its radius is $\xi$.
For the tetralayer graphene of interest~\cite{han_signatures_2025}, the situation is very different: as only one of the three components vanishes at the core of our TDs, we anticipate the effective viscosity $\eta_{\rm eff}$ for the TDs to be different from what one expects from the normal-state resistance. Furthermore, the core size need not be $\xi$ as it is for standard vortices. In the absence of microscopic details, it is therefore hard to predict the values of the resistance plateaus shown in Fig.~\figref{fig:general_picture}{b} and their duration. In fact, the underlying reason for the duration of the plateaus is hard to determine even in classic telegraph noise settings in nanostructures~\cite{ralls_individual-defect_1989}, and therefore we cannot make any claim regarding its origin in Ref.~\cite{han_signatures_2025}.

We can, however, propose a microscopic mechanism that accounts for the resistive plateaus observed in Ref.~\cite{han_signatures_2025} and illustrated in Fig.~\figref{fig:general_picture}{b}. Thermal fluctuations can cause weak spots in the sample to couple to the SC order and become sources of defect pairs. Once the TDs are created, they start moving due to the current, and generate resistance as per the Bardeen-Stephen theory~\cite{stephen_viscosity_1965} (see Sec.~SII of the Supplemental Material for elaboration). Their motion is along a vertical line [see Figs.~\figref{fig:defects_motion}{a--d}]. When a TD reaches the edge it disappears, since the lattice sites near the edge do not support 5-7 dislocations (their coordination number is lower than 6 anyway). However, the source keeps generating mobile TDs, and therefore the resistance maintains its steady-state value. The almost instantaneous onset of the resistive plateaus in Fig.~\figref{fig:general_picture}{b} suggests that the system quickly reaches the steady state, where $L_{y}/d$ defects are moving along the line. Indeed, for any reasonable estimate of the velocity of the TDs (see Sec.~SII of the Supplemental Material), the timescale for reaching the steady state is $<200\,{\rm ms}$, much shorter than the duration of the plateau (hundreds of seconds). Finally, the zero-resistance state can be restored by another random thermal event, which leads to a reconfiguration of the impurities in such a way that the defect source is turned off. When that happens, the remaining defects will make their way to the edge and quickly disappear, explaining the apparent abrupt vanishing of the resistive state.

 The dual identity of the TDs will manifest through extreme anisotropy in the above resistance resulting from the TD motion. The movement of dislocations in crystals requires bond switching events~\cite{stone_theoretical_1986} making it much easier for the dislocations to glide than climb. This aspect of crystalline defects gained new interest from the perspective of so-called  \emph{lineons}~\cite{gromov_colloquium_2024}: emergent particles whose motion is restricted to a lower-dimensional subspace of the space that supports the particle. Given that our TDs are defined atop an emergent ``crystal'' of PDW, the anisotropy may not be so severe as to prevent motion along one direction. Nonetheless, clearly distinct deformations required by the glide shown in Figs.~\figref{fig:defects_motion}{a--d} and climb shown in Figs.~\figref{fig:defects_motion}{f--g} would imply strong directional anisotropy in electrical transport that could provide an experimental signature of these exotic excitations.

\begin{figure}
    \centering
    \includegraphics[width=\linewidth]{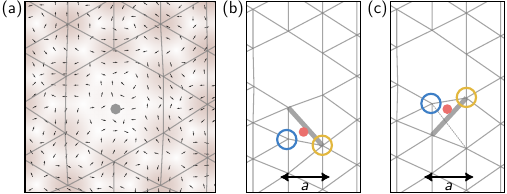}
    \caption{
    The effect of an external magnetic field $B_{\perp}$ on defects. (a) A field-induced vortex suppressing the pair amplitude and creating a vacancy, illustrated for the case of vortex creation at the local pair amplitude maximum. 
   (b-c)~A TD (red dot) associated with the 5-7 defect (blue and yellow circles) moves through a bond (thick grey line) switching among the lattice sites. A vacancy will present a roadblock for defects within a strip of width $a$ (the lattice separation).
    } 
    \label{fig:magnetic_field}
\end{figure}

We now turn our attention to the fate of the resistance fluctuations under an external out-of-plane magnetic field $B_{\perp}$.
$B_{\perp}$ will introduce external vortices. Each vortex will suppress the superconducting amplitude at its core and erase the emergent PDW ``latice site'' to introduce a vacancy, as illustrated for a simple case of a vortex landing on a local peak of superconducting pair amplitude in Fig.~\figref{fig:magnetic_field}{a}~
\footnote{A vortex landing at locations away from the emergent lattice site will suppress more lattice sites, albeit more weakly.}.
Since a TD moves by switching bonds [see Fig.~\figref{fig:magnetic_field}{b--c}],
introducing a vacancy through a field-induced vortex on the preferred path of TD's motion will hinder the TD motion. For the TD to move, it must circumvent the vacancy by departing from its preferred direction of movement. On the other hand, a vacancy at locations away from the preferred path of motion will not affect the TD motion. 
Assuming that the vortex has to land within the 
PDW lattice constant $a$ around the preferred path of TD, the likelihood that a vortex will disrupt the TD motion at low field is small if vortex could lend anywhere in the sample of width $L_x$. However, upon increasing the external field $B_{\perp}$ which introduces more and more full vortices, it will become more likely for one of those vortices to present a roadblock and stop the movement of TDs, halting the steady flow and generation of TDs. Hence the system will lose the telegraph noise like occurrence of resistive states at a threshold value of perpendicular magnetic field $B_\perp$. 

 For a ballpark estimate of the requisite field strength, we consider the external vortex to be equally likely to land anywhere on the sample. Then the probability of at least one of them being within the strip of width $a$ of the preferred path is $p=1-(1-a/L_{x})^{N_{\rm v}}$, where $N_{\rm v}=[ \Phi_{0}/(B_{\perp}L_{x}L_{y}) ]$ is the number of vortices ($[x]$ means rounding $x$ to the nearest integer). 
For the tetralayer graphene in Ref.~\cite{han_signatures_2025}, the resistance fluctuations disappear at $B_{\perp}\approx8\,{\rm mT}$, corresponding to $N_{\rm v}\approx108$ vortices (see Sec.~SII of the Supplemental Material). With the PDW lattice constant of $a\approx60\,{\rm nm}$ estimated in Ref.~\cite{gaggioli_spontaneous_2025} and $L_{x}=9.6\,\mu{\rm m}$, the blocking probability is roughly $p=0.1$ at the threshold field strength, which appears to be a reasonable likelihood for the blockage to occur. 

Taking a step back and analyzing our results, it is crucial to identify the main physical origins and consequences of the PDW order. The periodic, crystal-like structure formed by the PDW alone cannot explain the observed resistive behavior; despite containing many vortices and antivortices~\cite{gaggioli_spontaneous_2025}, these are not free to move and therefore cannot generate resistance. Since mobile vortices are the only plausible mechanism for finite resistance in a superconductor, the observation of such behavior at zero magnetic field is highly unusual. In standard superconductors, even when magnetic fields are present, pinning effects typically immobilize vortices. Therefore, the key requirement is a crystal whose defects have dual facets: they function as vortices, causing resistance when mobile, while simultaneously acting as crystalline defects that couple to charge disorder and occur naturally in two-dimensional crystals. This dual identity of the defects manifests the most salient property of PDWs: the coupling between positional modulation and the superconducting phase, which is conjugate to Cooper pair number. The multi-component PDW brings in precisely the necessary elements, and our subsequent analysis of resistive behavior follows from simple physical considerations. While we do not claim this is the only possible explanation for the observations reported in Ref.~\cite{han_signatures_2025}, reconciling superconductivity, resistive jumps, and a quarter-metal normal state through alternative mechanisms presents significant challenges.

Our theoretical analysis, combined with the experimental results~\cite{han_signatures_2025}, opens several avenues for future research. The restricted mobility of the defects could be directly tested experimentally by driving current in two perpendicular directions and comparing the resulting resistance values. We anticipate strong anisotropy in these measurements, as the triangular lattice structure inherently prevents both directions from being allowed.
Another prediction of our theory is that, if the current is not perpendicular to the Burgers vector but rather oriented at some angle $\theta$ with respect to it (here $\theta=\pm60^{\circ}$), then a Hall voltage will be induced due to the defects' motion, with Hall angle $R_{xy}/R_{xx}=\cot\theta$.
Such confirmation would position multi-component PDWs as the first genuine experimental realization of lineon dynamics in condensed matter systems. Moreover, the findings provide compelling motivation for further investigations into the properties of pure PDWs. Particularly intriguing are the nature of bound states in $\frac{1}{3}$ vortices and their implications for measurable properties, which remain largely unexplored. 

\emph{Acknowledgement.---}We are grateful to S.~A.~Kivelson and
M.~Rosales for insightful discussions. We thank L.~Ju and T.~Han for sharing their experimental data with us. 
We thank D.~F.~Agterberg, Y.~Wang, and R.~Thomale for useful comments on the manuscript.
This research was supported in part by grant NSF PHY-2309135 to the Kavli Institute for Theoretical Physics (KITP). O.L. is supported by a Bethe-KIC postdoctoral fellowship at Cornell University. O.L. and E.-A.K. are supported by the U.S. Department of Energy through Award Number DE-SC0023905. J.P.S. is supported by NSF DMR-2327094.

\bibliography{library}

%%%%%%%%%%%%%%%%%%%%%%%%%%%%%%%%%%%%%%%%%%%%%%%%%%
%%%% SUPPLEMENTAL MATERIAL %%%%%%%%%%%%%%%%%%%%%%%
%%%%%%%%%%%%%%%%%%%%%%%%%%%%%%%%%%%%%%%%%%%%%%%%%%

\clearpage
\setcounter{secnumdepth}{2}
\onecolumngrid

\begin{center}
\Large{\textbf{Supplemental Material}}
\end{center}

\setcounter{equation}{0}
\renewcommand{\theequation}{S\arabic{equation}}
\setcounter{figure}{0}
\renewcommand{\thefigure}{S\arabic{figure}}
\setcounter{section}{0}
\renewcommand{\thesection}{S\Roman{section}}

\section{Fractional vortices in a multi-component PDW}
Here we establish the existence of fractional vortices in the case of interest, which is a three-component PDW. To do so, we will adopt a Ginzburg-Landau (GL) treatment, along the lines of Ref.~\cite{rosales_electronic_2024}. We consider an $N$-component PDW order parameter (we will later take
$N=3$):
\begin{equation}
\Delta\left(\vec r\right)=\sum_{j=1}^{N}\Delta_{\vec G_{j}}\left(\vec r\right)e^{i\vec G_{j}\cdot\vec r}.
\end{equation}
The lowest-order terms allowed in the GL free energy are 
\begin{equation}
\begin{aligned}{\cal F}_{\text{PDW}}\left[\left\{ \Delta_{\vec G_{j}}\right\} _{j=1}^{N}\right] & =\kappa\sum_{j=1}^{N}\left|\nabla\Delta_{\vec G_{j}}\right|^{2}+r\sum_{j=1}^{N}\left|\Delta_{\vec G_{j}}\right|^{2} +u\left(\sum_{j=1}^{N}\left|\nabla\Delta_{\vec G_{j}}\right|^{2}\right)^{2}+\gamma\sum_{j\neq\ell}\left|\nabla\Delta_{\vec G_{j}}\right|^{2}\left|\nabla\Delta_{\vec G_{\ell}}\right|^{2}.
\end{aligned}
\end{equation}
We write each complex field $\Delta_{\vec G_{j}}$ as 
\begin{equation}
\Delta_{\vec G_{j}}\left(\vec r\right)=\left|\Delta_{\vec G_{j}}\left(\vec r\right)\right|\exp\left[i\theta_{\vec G_{j}}\left(\vec r\right)\right].
\end{equation}

We now consider fractional vortices. As the most basic example, we will assume that $\Delta_{\vec G_{1}}$ winds around the origin, whereas all $\left\{ \Delta_{\vec G_{j}}\right\} _{j=2}^{N}$ are constant.
We look for a configuration minimizing ${\cal F}_{\text{PDW}}$ subject to the long-distance behavior 
\begin{equation}
\begin{gathered}\lim_{\left|\vec r\right|\to\infty}\Delta_{\vec G_{1}}\left(\vec r\right)=\Delta_{\text{PDW}}e^{i\phi},\quad \lim_{\left|\vec r\right|\to\infty}\Delta_{\vec G_{j>1}}\left(\vec r\right)=\Delta_{\text{PDW}},
\end{gathered}
\end{equation}
where $\phi=\tan^{-1}\left(y/x\right)$ is the polar angle.
The assumption that at long distances all $\left\{ \left|\Delta_{\vec G_{j}}\right|\right\} _{j=1}^{N}$ are equal is a manifestation of our assumption that the PDW state itself does not break the $C_3$ symmetry.
Near the core of the defect, $\Delta_{\vec G_{1}}$ must go to zero whereas the others are uniform:
\begin{equation}
\begin{gathered}\lim_{\left|\vec r\right|\to0}\Delta_{\vec G_{1}}\left(\vec r\right)=0,\quad \lim_{\left|\vec r\right|\to0}\Delta_{\vec G_{j>1}}\left(\vec r\right)=\Delta_{0}.
\end{gathered}
\end{equation}
To find the explicit profile of the different order parameters, we take a non-linear sigma model approach~\cite{rosales_electronic_2024}. We define the $\left(N+1\right)$-component unit vector 
\begin{equation}
\vec n\left(\vec r\right)=\frac{1}{\Delta}\left(\text{Re}\Delta_{\vec G_{1}},\text{Im}\Delta_{\vec G_{1}},\Delta_{\vec G_{2}},\ldots\Delta_{\vec G_{N}}\right),
\end{equation}
with $\vec n^{2}\left(\vec r\right)=1$. Substituting it into ${\cal F}_{\text{PDW}}$, we obtain 
\begin{equation}
\begin{aligned}{\cal F}_{\text{PDW}} & =\kappa\left[\left(\nabla\text{Re}\Delta_{\vec G_{1}}\right)^{2}+\left(\nabla\text{Im}\Delta_{\vec G_{1}}\right)^{2}+\sum_{j>1}\left(\nabla\Delta_{\vec G_{j}}\right)^{2}\right]\\
 & +r\left[\left(\text{Re}\Delta_{\vec G_{1}}\right)^{2}+\left(\text{Im}\Delta_{\vec G_{1}}\right)^{2}+\sum_{j>1}\left(\Delta_{\vec G_{j}}\right)^{2}\right]\\
 & +u\left[\left(\text{Re}\Delta_{\vec G_{1}}\right)^{2}+\left(\text{Im}\Delta_{\vec G_{1}}\right)^{2}+\sum_{j>1}\left(\Delta_{\vec G_{j}}\right)^{2}\right]^{2}\\
 & +\gamma\left[\left(\left(\text{Re}\Delta_{\vec G_{1}}\right)^{2}+\left(\text{Im}\Delta_{\vec G_{1}}\right)^{2}\right)\sum_{j>1}\left(\Delta_{\vec G_{j}}\right)^{2}+\sum_{j,\ell>1}\left(\Delta_{\vec G_{j}}\right)^{2}\left(\Delta_{\vec G_{\ell}}\right)^{2}\right]\\
 & =\kappa\Delta^{2}\left(\nabla\vec n\right)^{2}+r\Delta^{2}\underbrace{\vec n^{2}}_{1}+u\Delta^{4}\underbrace{\left(\vec n^{2}\right)^{2}}_{1}+\gamma\Delta^{4}\left[\left(n_{1}^{2}+n_{2}^{2}\right)\sum_{a>2}n_{a}^{2}+\sum_{a,b>2}n_{a}^{2}n_{b}^{2}\right]\\
 & =\kappa\Delta^{2}\left(\nabla\vec n\right)^{2}+r\Delta^{2}+u\Delta^{4}+\gamma\Delta^{4}\left[\left(n_{1}^{2}+n_{2}^{2}\right)\sum_{a>2}n_{a}^{2}+\sum_{a,b>2}n_{a}^{2}n_{b}^{2}\right].
\end{aligned}
\end{equation}
In the uniform PDW phase (without the defect), the configuration is given by $\Delta_{\vec G_{j}}=\bar{\Delta}$ for all $j$, including $j=1$. In terms of $\vec n$, this translates into 
\begin{equation}
\vec n_{\text{PDW}}=\frac{1}{\sqrt{N}}\left(1,0,1,...,1\right).
\end{equation}
Substituting this into ${\cal F}_{\text{PDW}}$, we find 
\begin{equation}
\begin{aligned}{\cal F}_{\text{PDW}}\left(\vec n_{\text{PDW}}\right) & =r\bar{\Delta}^{2}+u\bar{\Delta}^{4}+\gamma\frac{1}{N^{2}}\begin{pmatrix}N\\
2
\end{pmatrix}\bar{\Delta}^{4} =r\bar{\Delta}^{2}+u\bar{\Delta}^{4}+\gamma\frac{N-1}{2N}\bar{\Delta}^{4}.
\end{aligned}
\end{equation}
The value of $\bar{\Delta}$ that minimizes the free energy is given
by
\begin{equation}
\begin{aligned}\bar{\Delta}^{2} & =-\frac{r}{2u+\gamma\frac{N-1}{N}}=-\frac{2Nr}{2Nu+\gamma\left(N-1\right)}.\end{aligned}
\end{equation}
Plugging this back into ${\cal F}_{\text{PDW}}$ for a general $\vec n$
vector, we find 
\begin{equation}
\begin{aligned}{\cal F}_{\text{PDW}} & =\kappa\bar{\Delta}^{2}\left(\nabla\vec n\right)^{2}+\gamma\bar{\Delta}^{4}\left[\left(n_{1}^{2}+n_{2}^{2}\right)\sum_{a>2}n_{a}^{2}+\sum_{a,b>2}n_{a}^{2}n_{b}^{2}\right]+\text{const.}\end{aligned}
\end{equation}

We would like to minimize this free energy subject to the boundary
conditions dictated by the defect. In terms of $\vec n$, they read
\begin{equation}
\begin{gathered}\lim_{\left|\vec r\right|\to\infty}\vec n\left(\vec r\right)=\frac{1}{\sqrt{N}}\left(\cos\phi\left(\vec r\right),\sin\phi\left(\vec r\right),1,\ldots,1\right),\\
\lim_{\left|\vec r\right|\to0}\vec n\left(\vec r\right)=\frac{1}{\sqrt{N-1}}\left(0,0,1,\ldots,1\right).
\end{gathered}
\end{equation}
Now we need to find a good parametrization of $\vec n$$\left(\vec r\right)$.
It makes sense to use polar coordinates along the ``$xy$'' plane,
defined by $n_{1}$ and $n_{2}$, and Cartesian coordinates for the
other dimensions. Then we can write 
\begin{equation}
\vec n\left(\vec r\right)=\sin\left[\alpha\left(\vec r\right)\right]\hat{\vec e}_{\vec r}+\cos\left[\alpha\left(\vec r\right)\right]\vec n_{\perp}\left(\vec r\right),
\end{equation}
where the unit vector $\vec n_{\perp}\left(\vec r\right)$ lies in
the $\left(N-1\right)$-dimensional hyper plane defined by $n_{3},\ldots,n_{N}$.
A reasonable assumption is
\begin{equation}
\vec n_{\perp}\left(\vec r\right)=\frac{1}{\sqrt{N-1}}\left(1,\ldots,1\right).
\end{equation}
The boundary conditions in terms of $\alpha$ are 
\begin{equation}
\begin{gathered}\lim_{\left|\vec r\right|\to\infty}\alpha\left(\vec r\right)=\sin^{-1}\left(\frac{1}{\sqrt{N}}\right),\\
\lim_{\left|\vec r\right|\to0}\alpha\left(\vec r\right)=0.
\end{gathered}
\end{equation}

We now plug the parametrized form into the free energy. The $\gamma$
term is simple: 
\begin{equation}
\begin{aligned}\left(n_{1}^{2}+n_{2}^{2}\right)\sum_{a>2}n_{a}^{2}+\sum_{a,b>2}n_{a}^{2}n_{b}^{2} & =\sin^{2}\alpha\cos^{2}\alpha+\cos^{2}\alpha\frac{1}{\left(N-1\right)^{2}}\begin{pmatrix}N-1\\
2
\end{pmatrix}\\
 & =\sin^{2}\alpha\cos^{2}\alpha+\frac{\cos^{2}\alpha}{2}\frac{N-2}{N-1}.
\end{aligned}
\end{equation}
The gradient term requires some more care, because $\hat{\vec e}_{\vec r}$
is itself $\vec r$ dependent:
\begin{equation}
\begin{aligned}\nabla\vec n & =\nabla\left(\sin\left[\alpha\left(\vec r\right)\right]\hat{\vec e}_{\vec r}\right)+\vec n_{\perp}\nabla\left(\cos\left[\alpha\left(\vec r\right)\right]\right)\\
 & =\hat{\vec e}_{\vec r}\cos\left[\alpha\left(\vec r\right)\right]\nabla\alpha\left(\vec r\right)+\left(\nabla\hat{\vec e}_{\vec r}\right)\sin\left[\alpha\left(\vec r\right)\right]-\sin\left[\alpha\left(\vec r\right)\right]\nabla\alpha\left(\vec r\right)\\
 & =\hat{\vec e}_{\vec r}\cos\left[\alpha\left(\vec r\right)\right]\nabla\alpha\left(\vec r\right)+\frac{1}{r}\hat{\vec e}_{\vec{\phi}}\sin\left[\alpha\left(\vec r\right)\right]-\vec n_{\perp}\sin\left[\alpha\left(\vec r\right)\right]\nabla\alpha\left(\vec r\right),
\end{aligned}
\end{equation}
where we have used $\nabla\hat{\vec e}_{\vec r}=\frac{1}{r}\hat{\vec e}_{\vec{\phi}}$.
The gradient square is then
\begin{equation}
\left(\nabla\vec n\right)^{2}=\left[\nabla\alpha\left(\vec r\right)\right]^{2}+\frac{1}{r^{2}}\sin^{2}\left[\alpha\left(\vec r\right)\right].
\end{equation}
We assume that $\alpha\left(\vec r\right)$ is isotropic, i.e., it
only depends on $r$, which simplifies the free energy to 
\begin{equation}
\begin{aligned}{\cal F}_{\text{PDW}}\left[\alpha\right] & =2\pi\int_{0}^{\infty}rdr\left[\bar{\kappa}\left(\frac{d\alpha}{dr}\right)^{2}+\bar{\kappa}\left(\frac{\sin\alpha}{r}\right)^{2}+\bar{\gamma}\cos^{2}\alpha\left(\sin^{2}\alpha+\frac{1}{2}\frac{N-2}{N-1}\right)\right],\end{aligned}
\end{equation}
where we defined $\bar{\kappa}\equiv\kappa\bar{\Delta}^{2}$, $\bar{\gamma}=\gamma\bar{\Delta}^{4}$.
Performing the change of variables $t=\ln\left(r/r_{0}\right)$, the $r^{2}$ factors cancel out in the first two terms. Since $r_{0}$
is arbitrary, we choose it conveniently to be $r_{0}^{2}=-\bar{\kappa}/\bar{\gamma}$
(the signs have to do with the original signs of $\kappa,\gamma$), leading to 
\begin{equation}
\begin{aligned}{\cal F}_{\text{PDW}}\left[\alpha\right] & =2\pi\bar{\kappa}\int_{-\infty}^{\infty}dt\left[\left(\frac{d\alpha}{dt}\right)^{2}+\sin^{2}\alpha-e^{2t}\cos^{2}\alpha\left(\sin^{2}\alpha+\frac{1}{2}\frac{N-2}{N-1}\right)\right].\end{aligned}
\end{equation}
The corresponding Euler-Lagrange equation is 
\begin{equation}
\frac{d^{2}\alpha}{dt^{2}}=\frac{1}{2}\left[1+\frac{1}{2}\frac{N-2}{N-1}e^{2t}-\cos\left(2\alpha\right)e^{2t}\right]\sin\left(2\alpha\right)
\end{equation}
subject to the boundary conditions 
\begin{equation}
\lim_{t\to-\infty}\alpha\left(t\right)=0,\ \lim_{t\to\infty}\alpha\left(t\right)=\sin^{-1}\frac{1}{\sqrt{N}}.
\end{equation}
While this equation is too complicated to be solved analytically, it can be solved numerically, yielding the order parameter profiles around the defect, and establishing the stability of the fractional vortex.

\section{Detailed resistance calculations with experimental data}
Here we provide the details of the numerical estimations made in the main text with regard to the experimental findings of Ref.~\cite{han_signatures_2025}. The dimensions of the sample are roughly $L_{x}=9.6\ \mu\text{m}$, $L_{y}=2.9\ \mu\text{m}$. This means that at $B_{\perp}=8\,{\rm mT}$, the number of external vortices is 
\begin{equation}
    N_{\rm v} = \left[\frac{\Phi_0}{B_{\perp}L_{x}L_{y}}\right] = 108
\end{equation}
(square brackets indicate rounding to the closest integer). Furthermore, the coherence length can be estimated from measurements of $H_{c2}$, the critical magnetic field where superconductivity is destroyed. For the state of interest (SC1), the estimate is $\xi\approx20\,{\rm nm}$. The normal-state resistance can be estimated by either increasing the temperature or going above $H_{c2}$, and it is $R_{\rm N}\approx2\,k\Omega$. The resistance fluctuations depicted in Fig.~(1b) of the main text are predominantly around the value $R_{s}\approx 45\,\Omega$.

We now outline the Bardeen-Stephen theory~\cite{stephen_viscosity_1965}, which describes the motion of vortices in a superconductor. We consider a two-dimensional superconductor of size $L_{x}\times L_{y}$ with vortices. Each vortex carries magnetic flux $\Phi$ (this will later be set to $\Phi_{0}$ for standard vortices, but for fractional vortices it is $\Phi_{0}/3$). A bias current is applied in the horizontal direction, with current per unit length $\vec J=J\hat{\vec x}$.

\subsection{Balance of forces on a single vortex}

A single vortex is a magnetic object with flux $\Phi$, corresponding to an out-of-plane magnetic field. The applied current causes a Lorentz force on the vortex, 
\begin{equation}
\vec F_{\text{L}}=\vec J\times\Phi\hat{\vec z}=J\Phi\hat{\vec y}.
\end{equation}
Phenomenologically, there is a drag force $\vec F_{\text{drag}}$ acting on the vortex, stopping it from accelerating indefinitely:
\begin{equation}
\vec F_{\text{drag}}=-\eta\vec v,
\end{equation}
where $\vec v$ is the velocity of the vortex, and $\eta$ is a phenomenological viscosity coefficient. By equating the forces, $\vec F_{\text{L}}=\vec F_{\text{drag}}$, we find the terminal velocity of the vortex: 
\begin{equation}
\vec v=\frac{\vec J\times\Phi\hat{\vec z}}{\eta}=\frac{J\Phi}{\eta}\hat{\vec y}.
\end{equation}

\subsection{Induced electric field due to vortex motion}

Consider now $N_{v}$ (identical) moving vortices. They all have the same velocity $\vec v$, because they all carry the same flux $\Phi$ and thus experience the same Lorentz and drag forces. The total flux they carry is $N_{v}\Phi$, and therefore the overall magnetic field associated with them is
\begin{equation}
\vec B_{\text{eff}}=\frac{N_{v}\Phi}{L_{x}L_{y}}\hat{\vec z}\equiv n_{v}\Phi\hat{\vec z},
\end{equation}
where we have defined the vortex density per unit area $n_{v}\equiv N_{v}/L_{x}L_{y}$. We now have moving vortices, so by Faraday's law, an electric field is induced: 
\begin{equation}
\vec E=-\vec v\times\vec B_{\text{eff}}=-\frac{J\Phi}{\eta}\hat{\vec y}\times n_{v}\Phi\hat{\vec z}=-\frac{n_{v}J\Phi^{2}}{\eta}\hat{\vec x}.\label{eq:induced_electric_field}
\end{equation}

\subsection{Resistance from induced electric field}

Ohm's law reads $\vec J=\sigma\vec E$, we $\sigma$ is the conductivity. The resistance in 2D is just the inverse of the conductivity, $R=1/\sigma$.
From Eq.~\eqref{eq:induced_electric_field} for the electric field, we find 
\begin{equation}
R=\frac{E}{J}=\frac{n_{v}\Phi^{2}}{\eta}=\frac{N_{v}\Phi^{2}}{\eta L_{x}L_{y}}.\label{eq:resistance_general}
\end{equation}
This is a general expression, and to make use of it we must find the viscosity $\eta$.

\subsection{Finding the viscosity}

To find the viscosity $\eta$, we consider the energy dissipation due to the vortex motion. Given the viscous force $-\eta\vec v$ and the velocity $\vec v$, the rate of energy dissipation is 
\begin{equation}
W=-\vec F_{\text{drag}}\cdot\vec v=\eta v^{2}.\label{eq:dissipation_eta}
\end{equation}
We now derive another expression for the dissipation, using a model for the vortex core. The simplifying assumption is that the core is a circle of radius $\xi$ in which the system is completely metallic,
with normal resistance $R_{\text{N}}$. According to the first London equation~\cite{tinkham_introduction_2004}, the electric field outside the core, induced by the motion of the vortex and the phase winding, is 
\begin{equation}
\vec E=\frac{\partial}{\partial t}\left(\frac{m^{*}}{e^{*}}\vec v_{s}\right)=-\vec v\cdot\nabla\left(\frac{m^{*}}{e^{*}}\vec v_{s}\right)=-\vec v\cdot\nabla\left(\frac{\hat{\vec{\theta}}}{2\pi\Phi_{0}r}\right).
\end{equation}
Here $m^{*}=2m_{e},e^{*}=2e$ are the effective mass and charge of the Cooper pair, $\vec v_{s}$ is the superfluid's velocity, and $\hat{\vec{\theta}}$ is a unit vector in polar coordinate. The term in parenthesis describe the circulation of the superconducting phase around the vortex. At $r>\xi$, this field is a dipole that averages out to zero. A nonzero average field can thus arise only from the core. The field inside the core is uniform, and can be found by continuity at $r=\xi$:
\begin{equation}
\vec E_{\text{core}}=\frac{v\Phi_{0}}{2\pi\xi^{2}}\hat{\vec y}.
\end{equation}
Using Ohm's law $\vec J_{\text{core}}=\vec E_{\text{core}}/R_{\text{N}}$, which is valid inside the normal core, we find the rate of energy dissipation in the core: 
\begin{equation}
W_{\text{core}}=\frac{\pi\xi^{2}E_{\text{core}}^{2}}{R_{\text{N}}}=\alpha\frac{v^{2}\Phi_{0}^{2}}{\xi^{2}R_{\text{N}}}.\label{eq:dissipation_core}
\end{equation}
We notice that several authors have proposed different dissipation mechanisms~\cite{tinkham_introduction_2004}, and therefore the numerical prefactor $\alpha\sim O\left(1\right)$ is debatable. Comparing the dissipation in Eq.~\eqref{eq:dissipation_eta} to that in Eq.~\eqref{eq:dissipation_core}, we find that 
\begin{equation}
\eta=\alpha\frac{\Phi_{0}^{2}}{\xi^{2}R_{\text{N}}}.
\end{equation}

\subsection{Estimating velocities}

We can estimate the velocity of a standard vortex $v_{0}$ from the above analysis. From the expression we found $v=I\Phi/\eta L_{x}$ and the expression for the viscosity $\eta$, we find that
\begin{equation}
v_{0}=\frac{I\Phi_{0}}{\eta L_{x}}=\frac{I}{L_{x}}\frac{R_{\text{N}}\xi^{2}}{\Phi_{0}}.
\end{equation}
In the experiment: 
\begin{equation}
\begin{aligned}R_{\text{N}}\approx2\ k\Omega & \quad\text{(measured at \ensuremath{T>T_{c}} or \ensuremath{H>H_{c}})}\\
L_{x}\approx9.6\ \mu\text{m} & \quad\text{(estimation of the sample's width)}\\
L_{y}\approx2.9\ \mu\text{m} & \quad\text{(estimation of the sample's length)}\\
\xi\approx20\:\text{nm} & \quad\text{(estimation in SC1 from \ensuremath{H_{c2}} measurements)}
\end{aligned}
\end{equation}
It is further reported that the bias current $I<0.5\ \text{nA}$. We can then estimate $v_{0}\approx2\times10^{-2}\ \text{m/s}$. A vortex therefore takes $t=L_{y}/v_{0}\approx145\,\mu\text{s}$ to cross the sample.

\subsection{Fractional vortices}

For fractional vortices, as explained in the main text, we cannot calculate the velocity or the resistance explicitly. The main reason is that their effective viscosity is strongly affected by the background PDW lattice, and their core is not entirely normal. Furthermore, the core radius is not necessarily $\xi$, but rather is likely related to the PDW lattice constant: the length scale over which the effect of the 5-7 topological defect is felt by the lattice. However, it is still reasonable to assume that the velocity of fractional vortices will not be radically different from that of standard vortices. Even if this estimate is off by three orders of magnitude, i.e., $t_{\text{defect}}\approx145\:\text{ms}$, it would
still appear as almost instantaneous at the scale of the duration of the resistive plateaus, which is hundreds of seconds. This explains the fast onset and turning off of the resistive fluctuations in the
experiment. 

In our picture of the flow of topological defects, the defect source generates them, the current makes them move, and when their distance from the source exceeds $d$, a new defect pair is generated. This leads to a steady flow of defects which gives a constant resistance. Over the length of the sample $L_y$, we have $N_{\rm defects}=L_y / d$ defects flowing in the steady state. While $d$ is not known, $d>a$ must hold, since a unit cell of the PDW is what defines the topological defect to begin with. Since $a>\xi$, we can safely assume $d>\xi$.

\end{document}